\documentclass[pre,superscriptaddress,twocolumn,a4paper,showpacs]{revtex4-1}
\usepackage{graphics}
\usepackage{amssymb}
\usepackage{wasysym}
\usepackage{latexsym}
\usepackage{graphicx}
\usepackage{epsfig}
\usepackage{subfigure}
\begin{document}

\title{Prediction diversity and selective attention in the wisdom of crowds}

\author{Davi A. Nobre}
\affiliation{Instituto de F\'{\i}sica de S\~ao Carlos,
  Universidade de S\~ao Paulo,
  Caixa Postal 369, 13560-970 S\~ao Carlos, S\~ao Paulo, Brazil} 
  
\author{Jos\'e F.  Fontanari}
\affiliation{Instituto de F\'{\i}sica de S\~ao Carlos,
  Universidade de S\~ao Paulo,
  Caixa Postal 369, 13560-970 S\~ao Carlos, S\~ao Paulo, Brazil}

% End title section

\begin{abstract}
The wisdom of crowds is the idea that the combination of independent estimates of the magnitude of some quantity  yields  a remarkably accurate prediction, which is always  more accurate  than the average  individual estimate.  In addition, it is largely believed that the accuracy of the crowd can be improved by increasing the diversity of the  estimates. Here we report the results of three experiments to probe the current understanding of the wisdom of crowds, namely, the   estimates of the  number of candies in a jar,   the length of a paper strip, and  the number of pages of a book.  
We find that the collective estimate is better than the majority of the  individual estimates in all three experiments. In disagreement with the  prediction diversity theorem,  we find no significant correlation between the prediction diversity and the collective  error.  The poor accuracy of the crowd on some experiments lead us to conjecture that its alleged accuracy   is most likely an artifice of selective attention.
\end{abstract}

%\begin{keywords}
%wisdom of crowds; forecast combinations; diversity prediction theorem; judgment distributions
%\end{keywords}

\maketitle

% The text of the paper follows. All of the text should be in the same file. 
% Use separate files for large tabular material and graphics.

\section{Introduction} \label{intro}

Although the notion of  wisdom of crowds is more than a century old, being brought to light by Galton's 1907 seminal study of the  estimation of the weight of an ox  at the West of England Fat Stock and Poultry Exhibition in Plymouth \cite{Galton_07} (see  \cite{Wallis_14} for an historical account), it is still a subject of  fascination  for the general public \cite{Surowiecki_04} and  for the scientific community \cite{Page_07} as well.   
This fascination  stems from  the  often remarkably accurate estimate given by the mean of independent individual estimates of the magnitude of some quantity.  For instance, in the ox-weighing contest,  the crowd overestimated the weight of the ox by less than 1\% of the correct weight \cite{Galton_07}.

Explanations for the accuracy of the collective estimate based on purely statistical arguments are not satisfactory, since they assume that the individual estimates are unbiased, that is, that the errors spread in equal proportion around the correct value of the unknown quantity \cite{Bates_69}, whereas experimental evidence, as well as common sense, points to the existence of systematic errors on the individual estimates \cite{Nash_14,Nash_17}. 

While the accuracy of the collective estimate remained a  sort of mystery, attention has been directed to the  observation  that  the  collective estimate  is always better than the average individual estimate. In fact, it seems that for some researchers this is the defining characteristic of the wisdom of crowds \cite{Mauboussin_07}. This observation can be explained by a simple
statistical argument, the so-called diversity prediction theorem  \cite{Page_07}, that asserts that the  error of the collective estimate is less  than or equal to the average error of the individual estimates.  Here, average  means simply the arithmetic mean of the individual estimates. 
This implies that, on the average, the collective estimate is better than the estimate of a  randomly selected individual in the group. This finding has considerable  practical importance as it guarantees that, in the event one does not  know who the experts are, it is advantageous to combine the forecasts of all members of the group. Nonetheless, the diversity prediction theorem offers no clue at all on the accuracy of the collective estimate.

However, as  hinted by its name,  the diversity prediction theorem seems also to have a say in the role of the diversity of the individual estimates, the so-called prediction diversity. In fact, since the theorem asserts  that the  (quadratic) collective error  equals the average (quadratic) individual error minus the diversity of the estimates, one is tempted to think that the increase of the prediction diversity would improve the collective estimate \cite{Page_07,Mauboussin_07}. Unfortunately, this result, which reflects somewhat  the zeitgeist of the 21st century, does not follow from the diversity prediction theorem since the average individual error and the  diversity of the estimates are not independent quantities, that is, an increase of the prediction diversity may change  the average  individual error with unpredictable effects on the collective error.  Nevertheless, unveiling the influence of the prediction diversity on the accuracy of the  collective estimate is clearly a crucial issue for  the understanding  of the wisdom of crowds.  

Accordingly, in this paper we report the results  of three experiments to probe the wisdom of crowds phenomenon, namely, the  classic estimate of the  number of candies in a jar,  the estimate of the length of a paper strip, and the estimate of the number of pages of a book.  The number of estimates in each experiment were over one hundred. To measure  the correlation between the prediction diversity and the collective error we produced a large number of virtual experiments by selecting $N$ estimates at random and without replacement  from the original ensemble of estimates of the real experiments. We find no significant correlation between the prediction diversity and the collective error, thus implying that diversity has no predictive value  for the accuracy of the collective estimate. Most interestingly, for the candies-in-a-jar and the pages-of-a-book experiments,  where the collective estimates  grossly missed the correct value,  there is an optimal group size that maximizes the chances of high-accuracy collective predictions, similarly to the findings on distributed cooperative  problem solving systems \cite{Fontanari_14}.  However, for the paper strip experiment, where the collective estimate  was already very accurate,  the chances of high-accuracy collective predictions (i.e., predictions that miss the correct value  by less than 5\%)  increase monotonically with  the group size $N$.

Regarding the accuracy of the collective estimate, our experiments support the view of systematic errors on the collective forecast \cite{Nash_14,Nash_17}, which depend on  the skills of the subjects on the proposed tasks.  For example, in the paper strip experiment the 
crowd underestimated the length by 1.8\% only, whereas in the pages-of-a-book experiment the underestimate was of 28.4\%. 
Our conclusion is that the high accuracy of the wisdom of crowds, which is responsible  for its popularity among the general public,  is an illusion resulting  from selective attention that gives prominence to the successful  outcomes only.

The rest of this paper is organized as follows. In Section \ref{sec:DPT} we  present a brief review of  the diversity prediction theorem \cite{Page_07} and introduce the basic quantities  used to characterize the wisdom of crowds experiments.  In Section \ref{sec:exp}
  we describe  and analyze the results of our three experiments, emphasizing the influence of the prediction diversity on the 
accuracy of the collective estimate.  Finally, in Section \ref{sec:disc}  we summarize our results and present  our concluding remarks.

%%%%%%%%%%%%%%%%%%%

\section{Page's Diversity Prediction Theorem}\label{sec:DPT}

The diversity prediction theorem \cite{Page_07} is considered a  main attainment to  those that celebrate  the power of diversity to improve the performance of groups \cite{Mauboussin_07} (see, however,  \cite{Fontanari_16}), since it  shows that the (quadratic) collective error  can be related in a very simple manner to the average  (quadratic) individual error  and to a measure of the diversity of the estimates. More pointedly, let $g_i$ be the estimate of some unknown quantity, such as the weight of the ox in Galton's experiment,  by individual $i=1, \ldots, N$. We will consider the $g_i$s as random variables that, as far as the diversity prediction theorem concerns, need not be independent. In addition, let $G$ be the true value of the unknown quantity the $N$ individuals are trying to estimate and the collective estimate be defined as the arithmetic mean of the individual estimates, that is, 
$ \langle g \rangle = \sum_{i=1}^N g_i/N$.  (We note that Galton used the median of the individual estimates as the collective estimate  in the ox-weighing experiment \cite{Galton_07},  though the arithmetic mean proved to be a better estimator in that case \cite{Wallis_14}.)
Thus,   the  quadratic collective error is defined as 
\begin{equation}\label{gamma}
\gamma =  \left ( \langle g \rangle - G \right )^2 .
\end{equation}
Next we define the average quadratic individual error,
\begin{equation}\label{eps}
\epsilon = \frac{1}{N} \sum_{i=1}^N \left ( g_i - G \right )^2,
\end{equation}
and the diversity of the estimates,
\begin{equation}\label{delta}
\delta = \frac{1}{N} \sum_{i=1}^N \left ( g_i  - \langle g \rangle  \right )^2,
\end{equation}
so that the identity
\begin{equation}\label{DPT}
\gamma =  \epsilon - \delta
\end{equation}
follows straightforwardly.
This  identity is Page's diversity prediction theorem, which  asserts  that the  (quadratic) collective error  equals the average (quadratic) individual error minus the prediction diversity.  This result is sometimes  interpreted as the proof that increasing the prediction diversity $\delta$ results in the decrease of the collective error $\gamma$ \cite{Mauboussin_07}. Of course, since $\delta$ and $\epsilon$ cannot be varied independently of each other,  this interpretation is not correct.  %For instance, the optimal scenario $g_i = G$ for all $i$  yields $\epsilon = \delta = 0$. 

In fact, the  discussion on the value of Page's diversity prediction theorem is reminiscent of the arguments about the relevance of the celebrated Price equation \cite{Price_70} for evolutionary biology. We note that  Price's equation, which has a straightforward derivation from the definition of fitness,  is considered by many researchers  as a mere mathematical tautology  \cite{Frank_12}. 

Here we take a pragmatic stance and carry out experiments to check whether the increase of the  diversity of the estimates  is likely to result in a decrease of the collective error, without entering into the merit of the diversity prediction theorem. It is interesting to note that the diversity of estimates $\delta$ is known in the statistical literature as the precision of the estimates, that is, the closeness of repeated estimates (of the same quantity)  to one another \cite{Tan_14}.
The experiments and the analyses of their results  are the subjects of the next section.

%%%%%%%%%%%%%%%%%%%

\section{Experiments}\label{sec:exp}

We have carried out three experiments in which a number of STEM  (science, technology, engineering, and mathematics) students of the University of S\~ao Paulo  guessed independently the number of candies in a jar, the length of a paper strip and  the number of pages of a book.  
%Next we describe and analyze the results of  these experiments.

\subsection{Estimating the number of candies in a jar}

This is a classic wisdom of crowds experiment  \cite{Surowiecki_04}  in which the individuals have to guess independently  the number of candies in a jar (see \cite{King_11} for a variant with non-independent guesses) resulting, typically, in a  group estimate superior to the vast majority of the individual guesses \cite{Mauboussin_07}. In particular, 105 students  guessed the number of candies in a transparent jar that held $G=636$ candies. The group estimate $\langle g \rangle =  531$  was better than 70\% of the individual estimates. In our setup, there was no penalty for the guess farthest from the correct answer and the reward for the best guess (630, in our case)   was the jar of candies.

\begin{figure}[t]
\centering  
\includegraphics[width=0.48\textwidth]{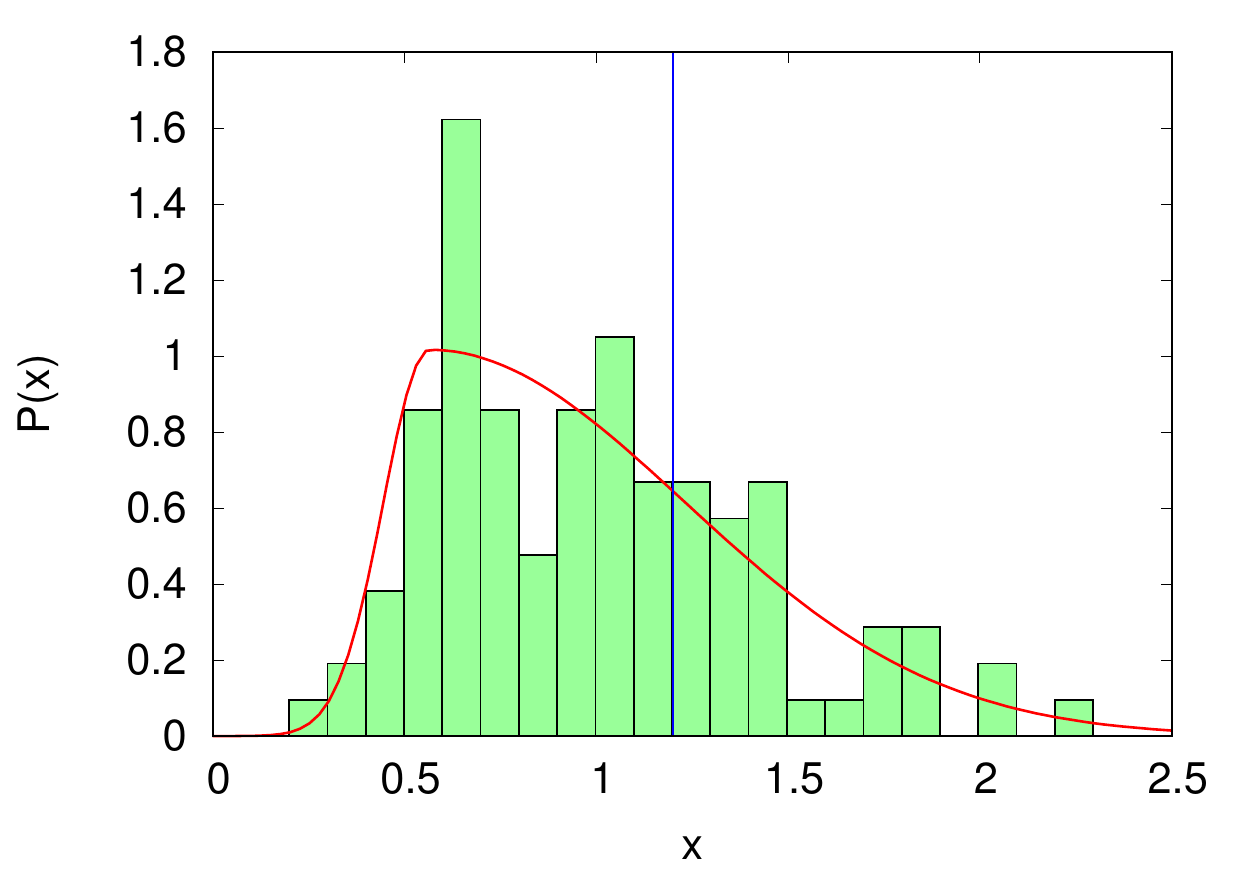} 
\caption{Histogram of the normalized estimates $x = g_i/ \langle g \rangle$ for the candies-in-a-jar experiment. The vertical line indicates the ratio between the correct number of candies $G =636$  and the estimate of the group $\langle g \rangle =  531$, that is,  $G/\langle g \rangle \approx 1.2$. The solid curve is the best fitting ($R^2 = 0.77$) two-pieces normal distribution $ A \,  e^{-\left(x-\mu \right)^2/2\sigma_1^2}$  if $x < \mu$ and $ A \, e^{-\left(x-\mu \right)^2/2\sigma_2^2} $ otherwise, where $A = \left [ \sqrt{2 \pi} \left ( \sigma_1 + \sigma_2 \right )/2 \right ]^{-1}$
with   $\sigma_1 = 0.12$,  $\sigma_2 = 0.66$.  and $\mu = 1 - (\sigma_2 -\sigma_1) \sqrt{2/\pi} \approx  0.565$ so that the mean of the fitting distribution equals 1.  }
\label{fig:1}
\end{figure}

Figure \ref{fig:1} shows the histogram of the  normalized guesses $g_i/\langle g \rangle $ which we attempted to fit using a two-pieces normal distribution  of mean 1 (see, e.g., \cite{Wallis_14b}).  Notice that the most probable estimate  is about half of the correct number  of candies in the jar.  The asymmetry of the distribution of estimates is quite noticeable and  is confirmed by the  positive value of the moment coefficient of skewness $\tilde{\mu}_3 = 0.73$. This means that  the odds of gross overestimation of the outcome are much greater than of underestimation due to the right-tailed nature of the distribution. It is interesting that Galton's  ox-weighing experiment  results in  a left-skewed, left-tailed  distribution of estimates  \cite{Wallis_14}. In fact, in order to tackle the asymmetry of the distribution, Galton  suggested the use of two normal distributions to fit the lower and the upper halves of the histogram of estimates, hence our choice of the two-pieces normal distribution as the fitting distribution in Figure \ref{fig:1}.

%----------------------------------------------------------------------------------------------------
%----------------------------------------------------------------------------------------------------

\begin{figure*}
\centering  
\includegraphics[width=0.38\textwidth]{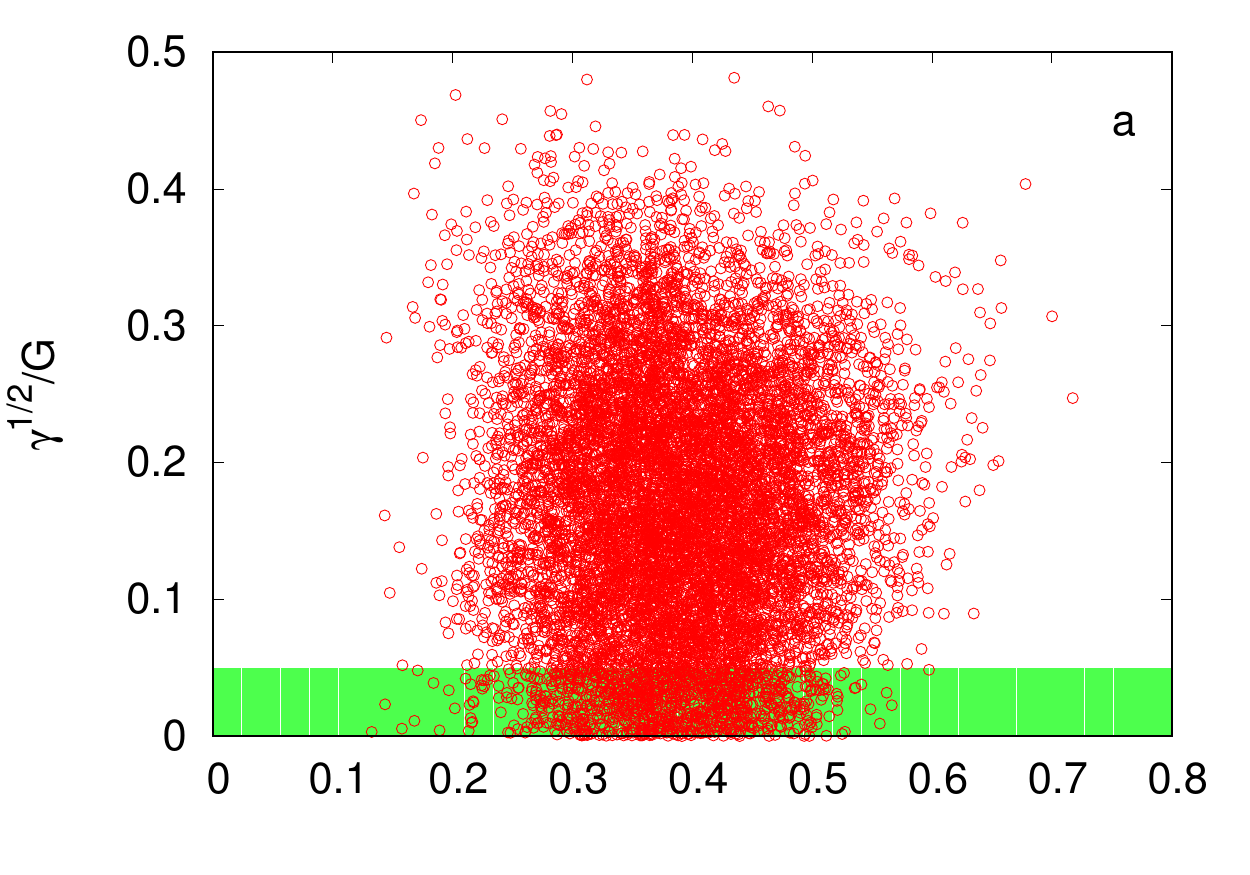} 
\includegraphics[width=0.38\textwidth]{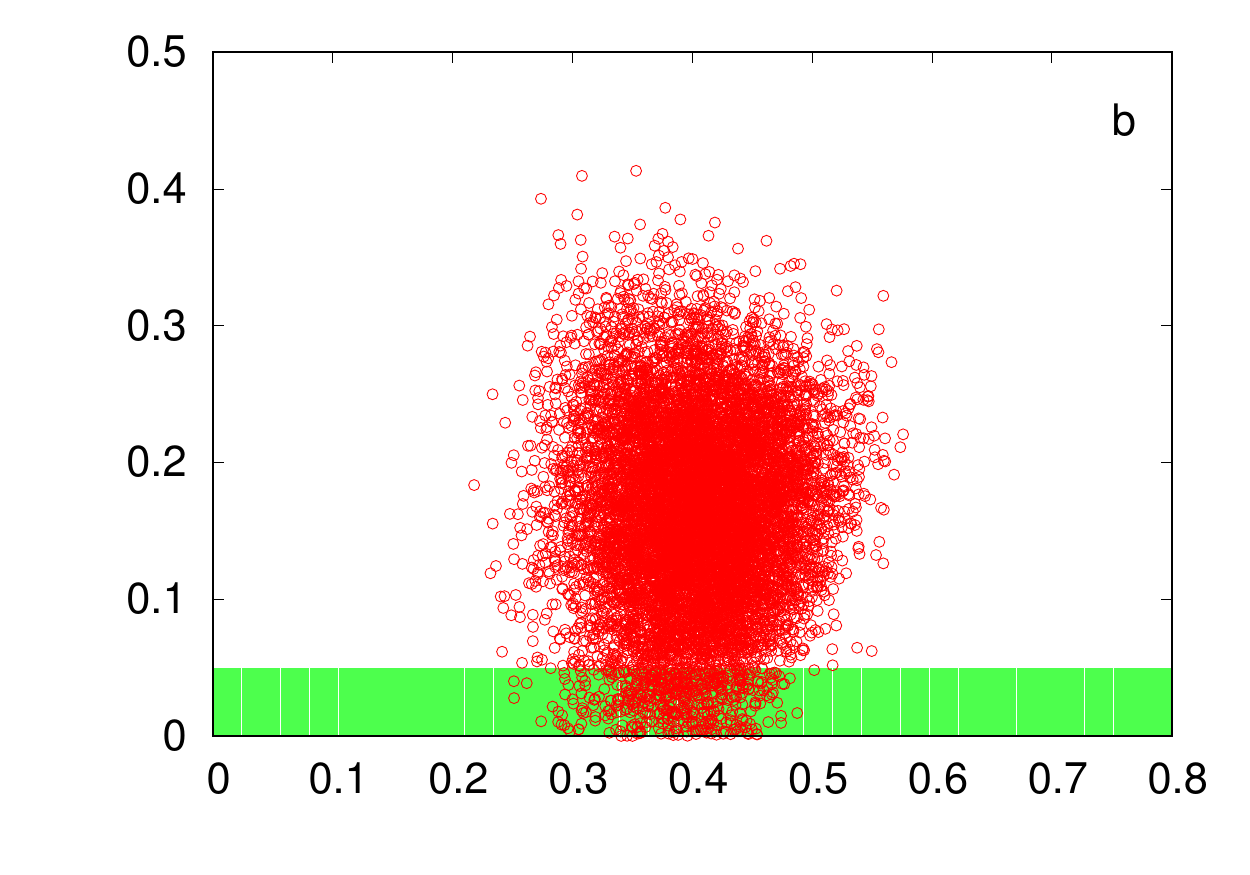} \\
\includegraphics[width=0.38\textwidth]{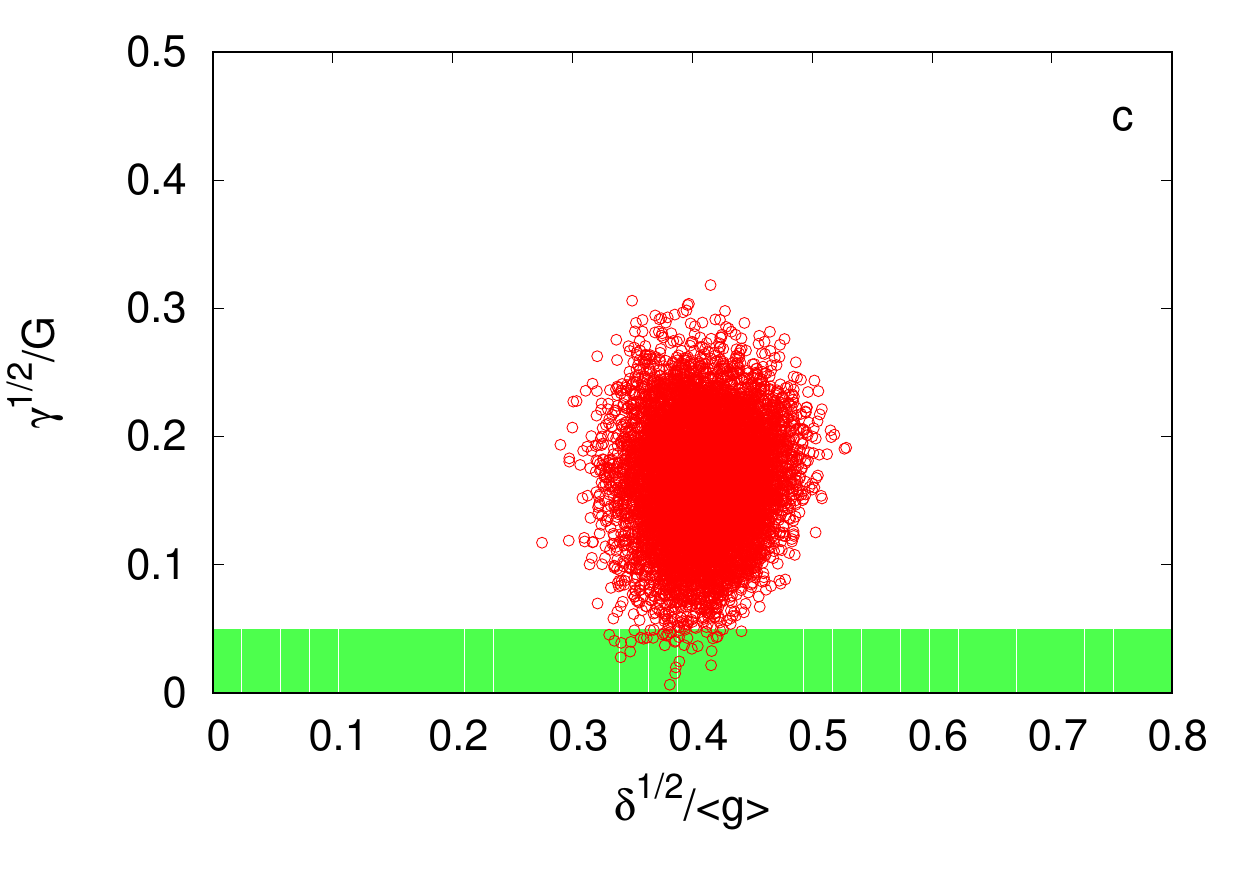} 
\includegraphics[width=0.38\textwidth]{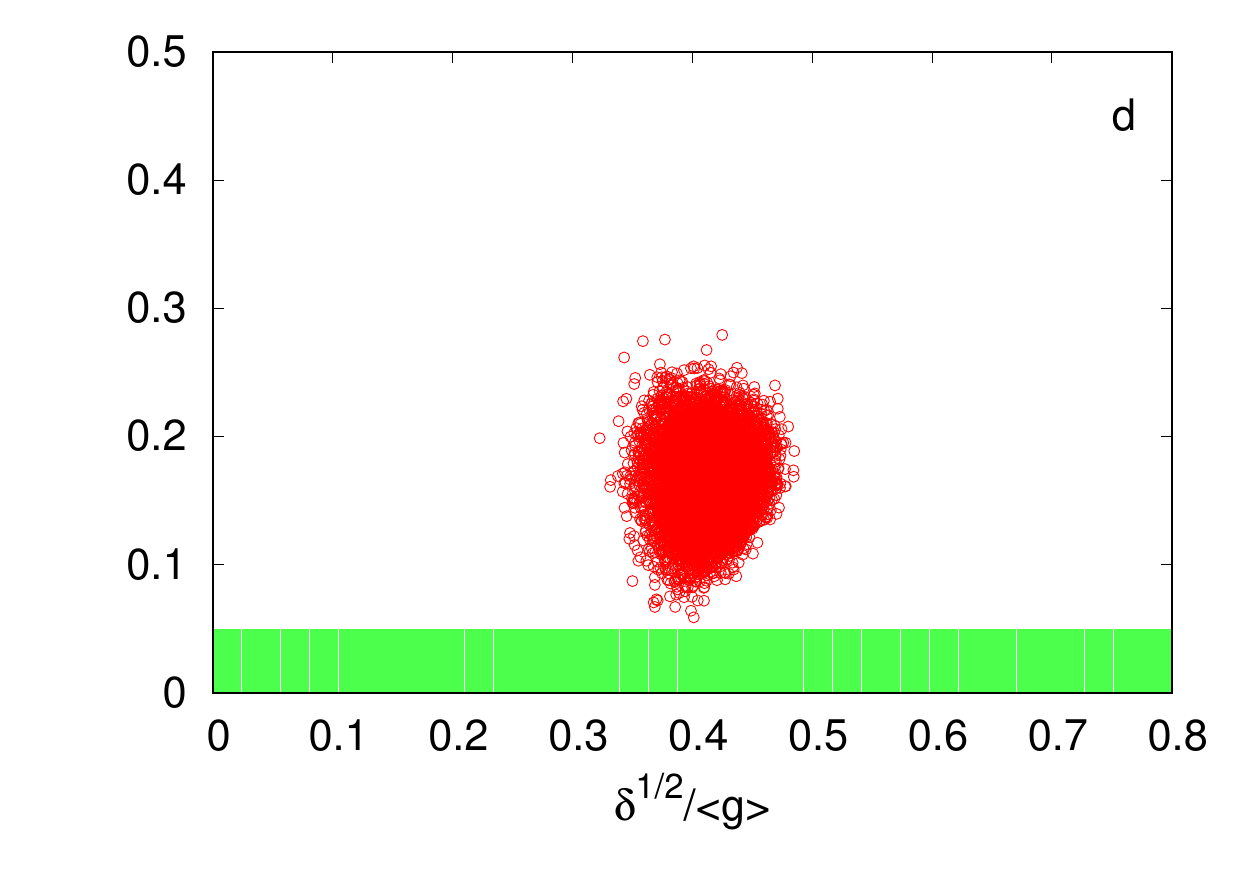}
\caption{Scatter plots of the candies-in-a-jar experiment for groups of  size a: $N=10$, b: $N=20$, c: $N=40$, d: $N=60$. The x-axis is the relative standard deviation of the estimates $\delta^{1/2}/\langle g \rangle $, which measures the diversity of the estimates, whereas the y-axis is the relative collective error $\gamma^{1/2}/G$, which measures the accuracy of the group estimate.
The  horizontal band at the bottom of the panels indicate the regions where  the percent error of the group estimate is less than 5\%.   }
\label{fig:2}
\end{figure*}

Since for a single experiment,  we cannot drawn any conclusion about the influence of the diversity  of the estimates $\delta$ on the collective  error $\gamma$, here we use the  estimates of the candies-in-a-jar experiment  to generate $10^4$  virtual experiments. In each experiment,   $N $ estimates are drawn at random without replacement from the $105$  original estimates. For each experiment,  $\gamma$ and $\delta$ are evaluated so we can draw the  scatter plots shown in Figure \ref{fig:2}. In particular, this figure shows the relative collective error $\gamma^{1/2}/G$ and the relative diversity $\delta^{1/2}/\langle g \rangle$ for each one of the $10^4$ experiments for groups of size $N=10,20,40$ and $60$. We note that the mean $\langle g \rangle$ depends on the particular experiment considered, so it assumes a different value for each point in the scatter plots. In terms of these dimensionless quantities, equation (\ref{DPT}) is rewriten as
\begin{equation}\label{DPT_2}
\frac{\gamma}{G^2} =  \frac{\epsilon}{G^2} -  \frac{\delta}{\langle g \rangle^2}  \frac{\langle g \rangle^2}{G^2} ,
\end{equation}
which preserves the main points of the diversity prediction theorem, namely, the correct claim that $\gamma^{1/2}/G \geq  \epsilon^{1/2}/G$  and the incorrect inference that the increase of the relative diversity $\delta^{1/2}/\langle g \rangle$ implies a decrease of  the relative collective error  $\gamma^{1/2}/G$. We hasten to note that although, in principle, these two quantities might be negatively correlated, this result does not follow from equation  (\ref{DPT_2}) since the  increase of $\delta$ may result in a decrease of  $\epsilon$ and $\langle g \rangle$.

%---------------------------------------------------------------------------------------------------- 
%----------------------------------------------------------------------------------------------------

%----------------------------------------------------------------------------------------------------
\begin{figure}
\centering  
\includegraphics[width=0.48\textwidth]{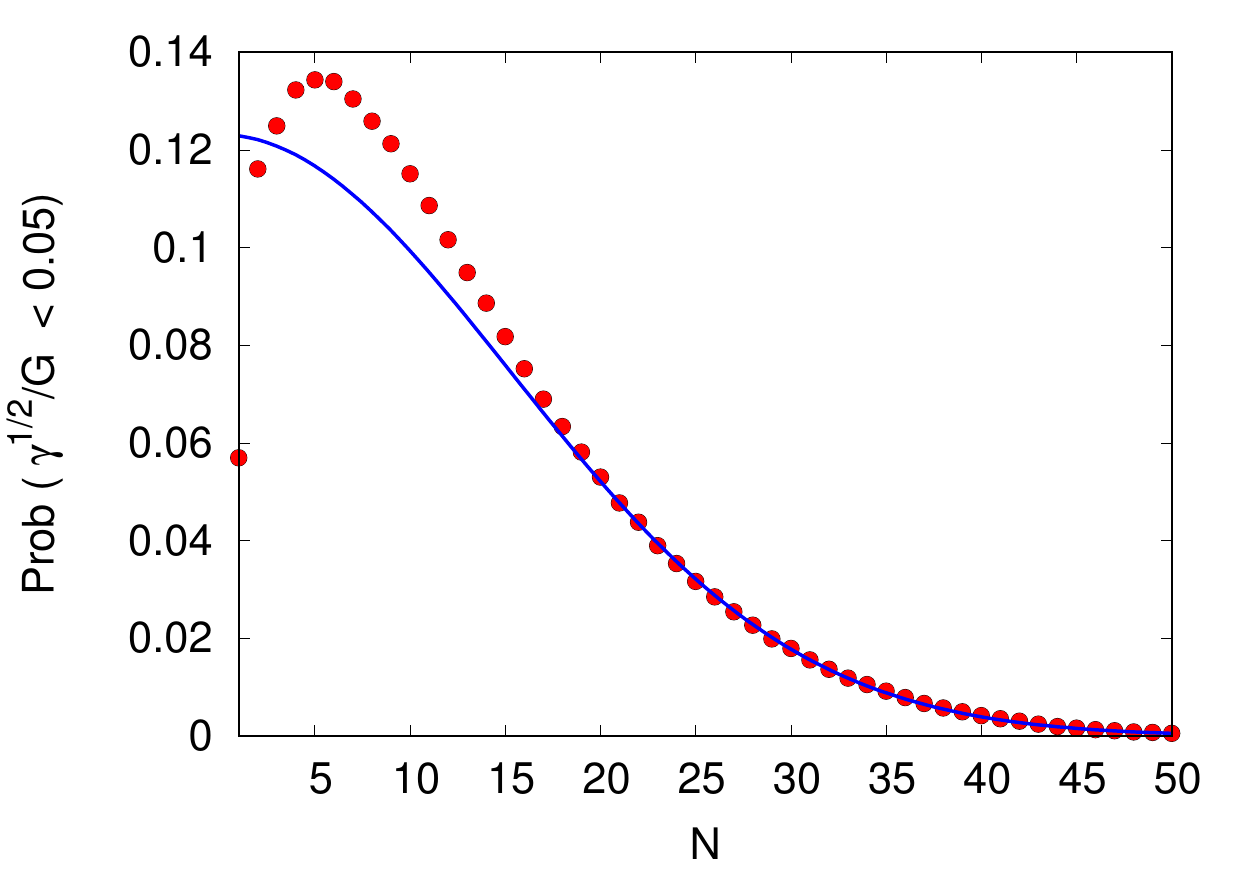} 
\caption{Probability that the percent collective error is less than 5\% for the candies-in-a-jar experiment  as function of the group size $N$. The solid curve is the fitting function $\alpha e^{-\beta N^2} $ for the large $N$ regime. The fitting parameters are
 $\alpha = 0.12$ and $\beta = 0.0021$. }
\label{fig:3}
\end{figure}
%----------------------------------------------------------------------------------------------------

%----------------------------------------------------------------------------------------------------
\begin{figure}[h]
\centering  
\includegraphics[width=0.48\textwidth]{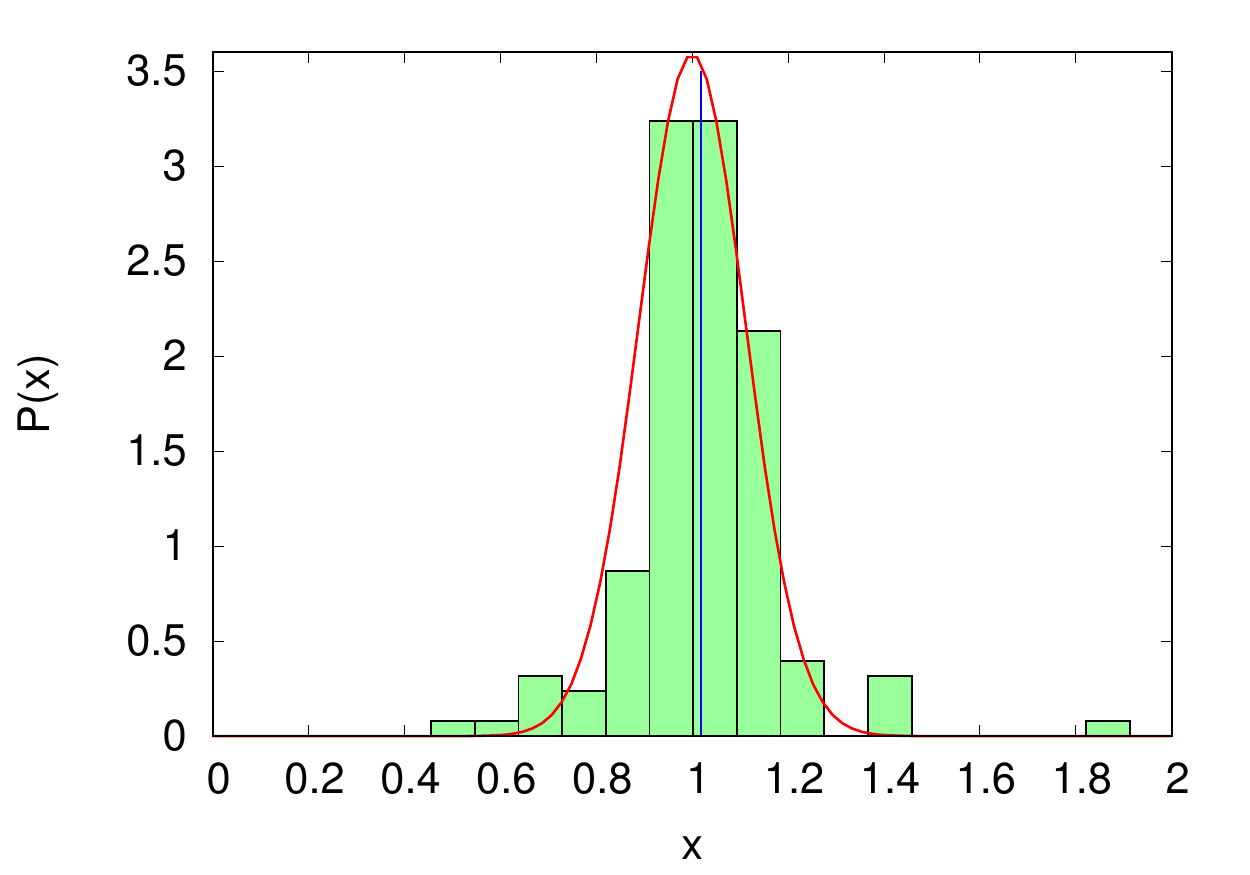} 
\caption{Histogram of the normalized estimates $x = g_i/ \langle g \rangle$ for the paper strip experiment. The vertical line indicates the ratio between the length of the paper strip $G =22.4~ \mbox{cm}$  and the estimate of the group $\langle g \rangle =  22.0~ \mbox{cm}$, that is,  $G/\langle g \rangle \approx 1.018$. The solid curve is the best fitting ($R^2 = 0.94$) Gaussian distribution $ e^{-\left(x-1 \right)^2/2\sigma^2} /\sqrt{2 \pi \sigma^2}$
with $\sigma^2 = 0.012$ and mean 1.  }
\label{fig:4}
\end{figure}
%----------------------------------------------------------------------------------------------------

The Pearson correlation coefficients between  $\delta^{1/2}/\langle g \rangle$ and  $\gamma^{1/2}/G$  are $r= -0.005$ for $N=10$, $r = 0.04$ for $N=20$, $r = 0.06$ for $N=40$ and $r = 0.07$ for $N=60$ so it is safe to state that  the diversity of the estimates conveys very little, if any, information on the accuracy of the group estimate. We note that the center of mass of the data shown in the panels of Figure \ref{fig:2} are at $\delta^{1/2}/\langle g \rangle = 0.416$ and  $\gamma^{1/2}/G = 0.165$ regardless of the value of $N$. Of course, these values coincide with those calculated using the original sample of 105 estimates.

An interesting result  revealed in Figure \ref{fig:2}  is that the probability that the percent error of the group estimate is less than 5\%, which is given by the fraction of data points  that fall within  the horizontal band in the scatter plot, depends on the group size $N$. To quantify this effect, we present in Figure \ref{fig:3} this probability as function of the group size. These startling  results  reveal the existence of an optimal group size ($N=5$) that maximizes the odds of producing a high accuracy estimate of the number of candies in the jar.  In order words,  picking $N=5$ estimates at random out of the 105 original ones yields a 14\% chance of  producing a group estimate that misses the correct value  by less than 5\%.  In addition, selection of a single estimate at random  is more likely to result in such high accuracy estimates than the aggregation of $N \geq 20$ estimates. The chance of such a precise estimate vanishes like $e^{-\beta N^2} $ with $\beta > 0$ for increasing $N$. This is so because the percent collective error of the original sample of 105 estimates, namely, $16.5 \%$, does not tally with  our definition of high accuracy  estimate.

%These results are reminiscent of Condorcet's jury  theorem

\subsection{Estimating the length of a paper strip}

Whereas the competence of the students to estimate the  number of candies in a jar is very problematical, as our experiment was their first experience on that task, we expect them to be better  skilled to size up  the length of a paper strip.  Accordingly, we asked 139 students  to guess the length of a paper strip that measured $G=22.4~ \mbox{cm}$. The group estimate $\langle g \rangle =  22.0~ \mbox{cm}$  was better than 85\% of the individual estimates and corresponds to a percent error of only $1.8 \%$. The best guess was $22.5~ \mbox{cm}$.
There was no reward or penalty for the subjects in this  experiment.

%---------------------------------------------------------------------------------------------------- 
\begin{figure*}[t]
\centering  
\includegraphics[width=0.38\textwidth]{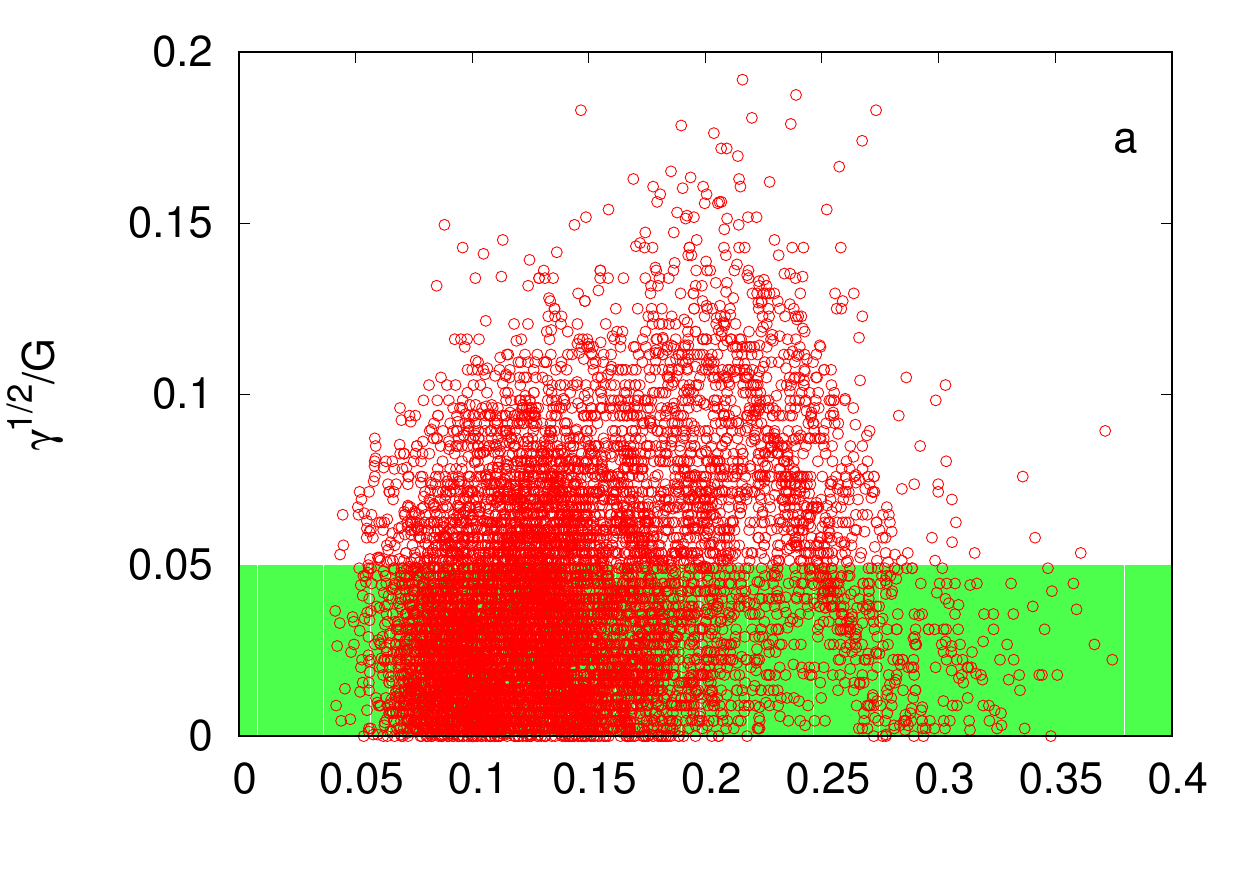} 
\includegraphics[width=0.38\textwidth]{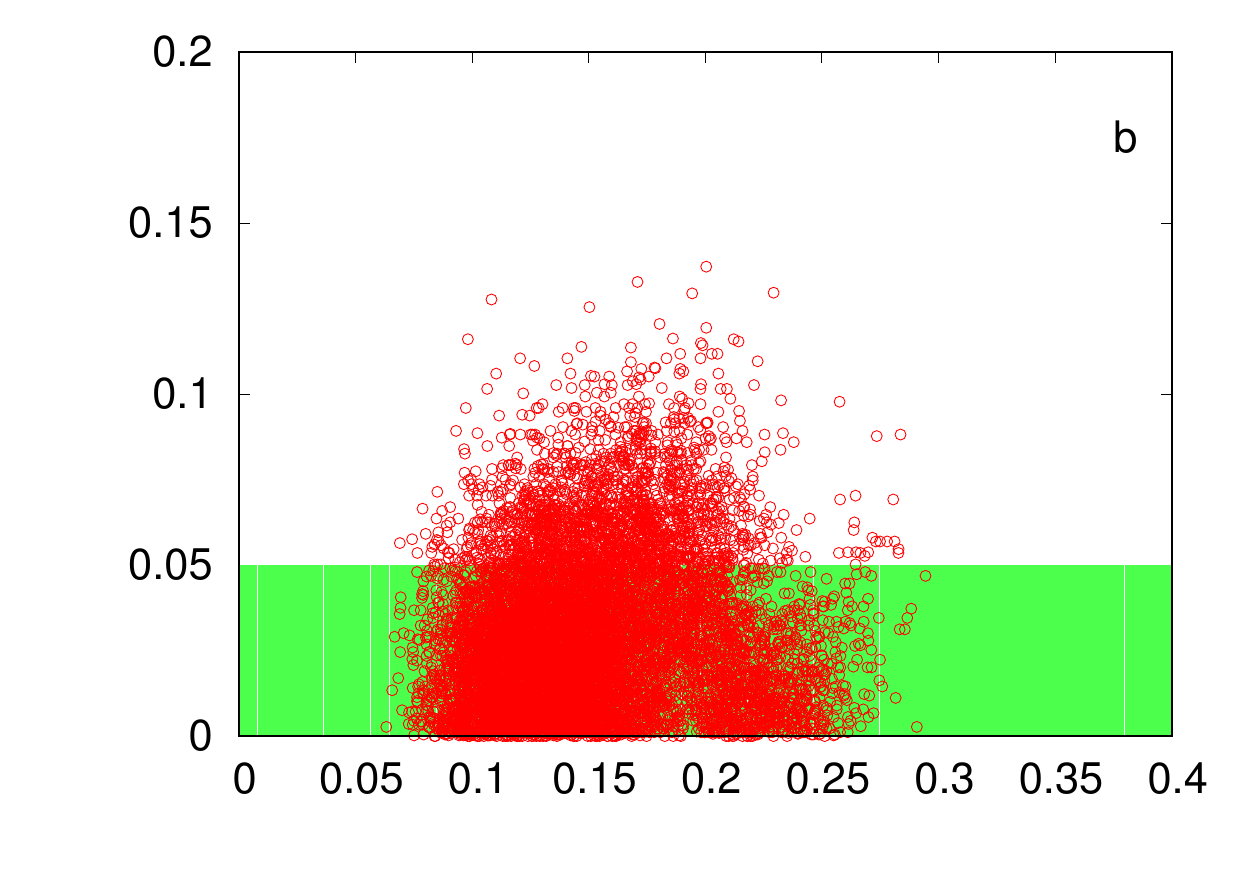} \\
\includegraphics[width=0.38\textwidth]{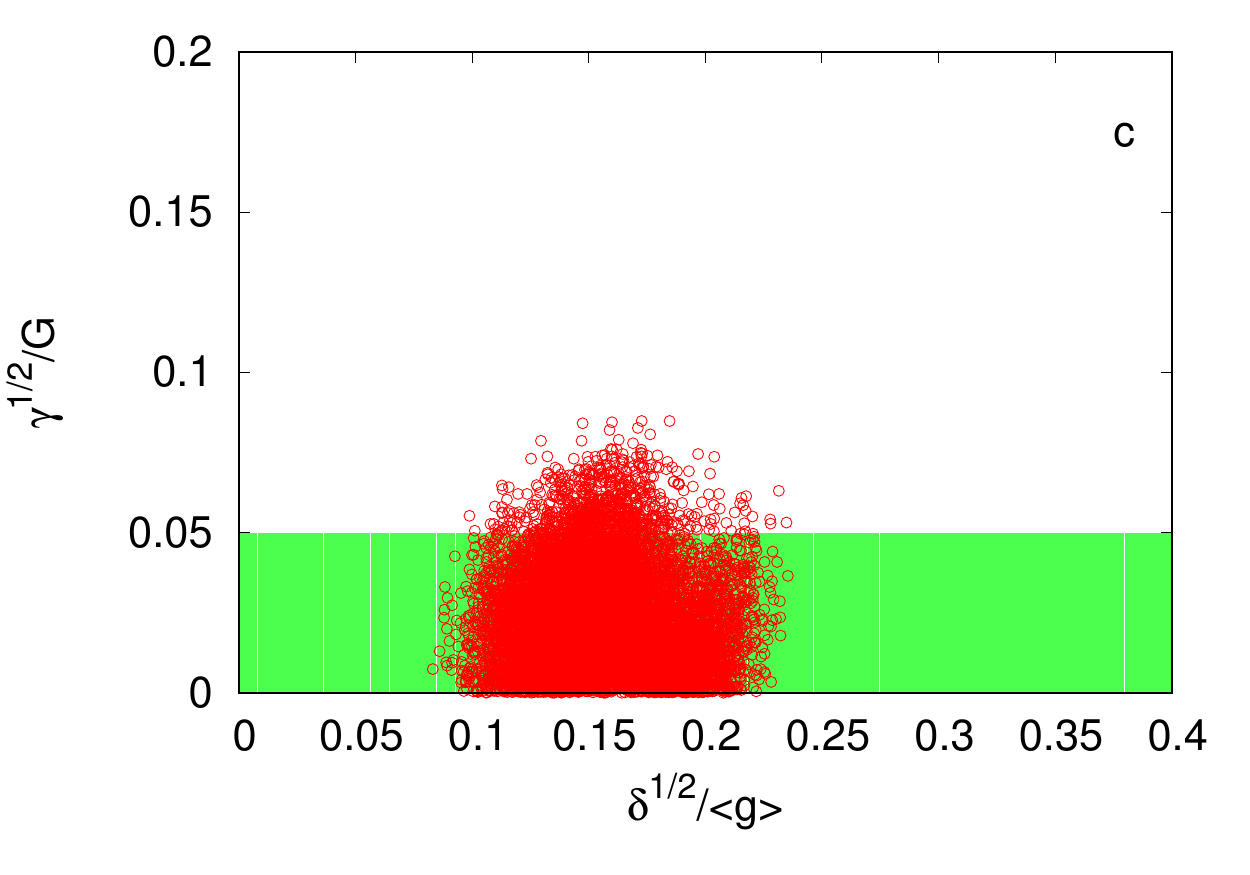} 
\includegraphics[width=0.38\textwidth]{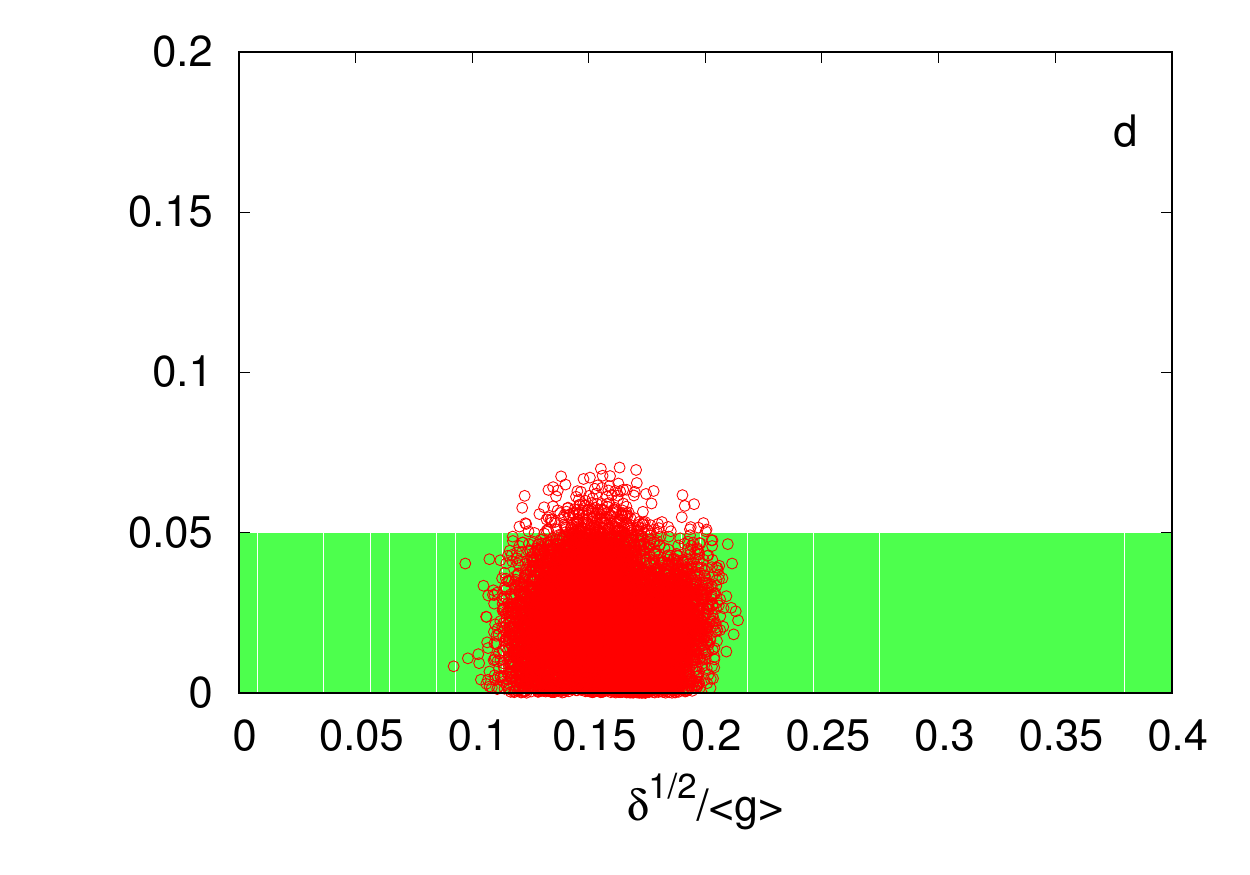}
\caption{Scatter plots of the paper strip experiment for groups of  size a: $N=10$, b: $N=20$, c: $N=40$, d: $N=60$. The x-axis is the relative standard deviation of the estimates $\delta^{1/2}/\langle g \rangle $, which measures the diversity of the estimates, whereas the y-axis is relative collective error $\gamma^{1/2}/G$. which measures the accuracy of the group estimate.
The  horizontal band at the bottom of the panels indicate the regions where  the percent error of the group estimate is less than 5\%.   }
\label{fig:5}
\end{figure*}
%----------------------------------------------------------------------------------------------------

The histogram of the  normalized guesses $x = g_i/\langle g \rangle $ for the paper strip experiment is shown in Figure \ref{fig:4} together with a best fitting Gaussian distribution. The symmetry of the estimates around the mean  and their small variance  attest the skill of the students to size up ordinary lengths. Although the noticeable asymmetry of the histogram of estimates for the candies-in-a-jar experiment  leads us to follow Galton's suggestion and use a two-pieces normal distribution as the fitting distribution (see Figure \ref{fig:1}), a normal distribution seems way more suitable to fit the almost symmetric histogram of the paper strip experiment.

\begin{figure}
\centering  
\includegraphics[width=0.48\textwidth]{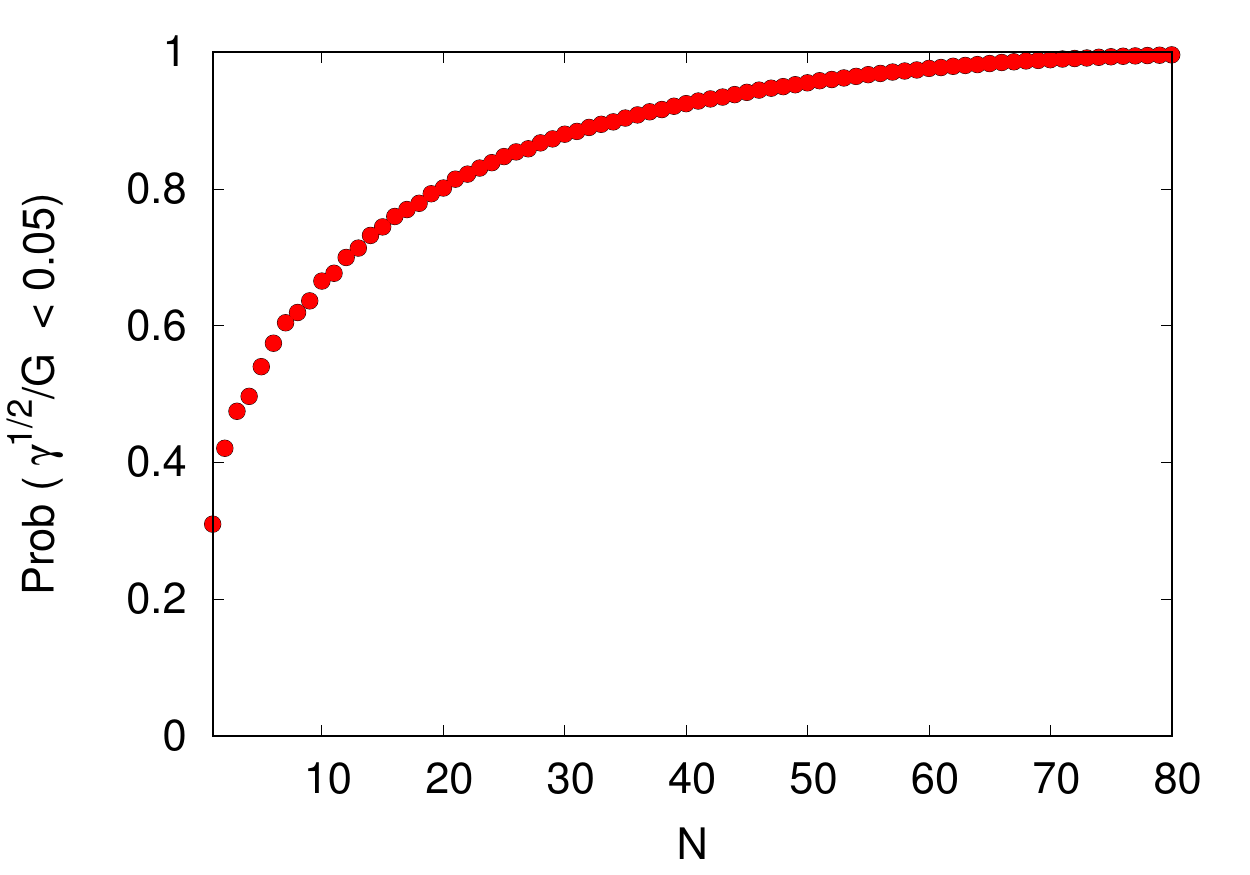} 
\caption{Probability that the percent collective error is less than 5\% for the paper strip experiment  as function of the group size $N$.  }
\label{fig:6}
\end{figure}

The scatter plots of Figure \ref{fig:5} that show  the properly normalized  diversity of the estimates and the relative collective error for distinct group sizes reveal  no meaningful correlation between these quantities. More pointedly,
the Pearson correlation coefficients between  $\delta^{1/2}/\langle g \rangle$ and  $\gamma^{1/2}/G$  are $r= 0.28$ for $N=10$, $r = 0.11$ for $N=20$, $r =- 0.049$ for $N=40$ and $r = -0.11$ for $N=60$. We note that the claim that high prediction diversity leads or, more precisely, is associated  to lower collective errors should be supported by a large negative correlation between $\delta^{1/2}/\langle g \rangle $ and $\gamma^{1/2}/G$. Although this correlation becomes more negative as $N$ increases, it is too low to offer useful information  on  the wisdom of crowds puzzle. 
The center of mass of the data shown in the panels of Figure \ref{fig:5} are at $\delta^{1/2}/\langle g \rangle = 0.16$ and  $\gamma^{1/2}/G = 0.018$ regardless of the value of $N$. The fraction of group estimates whose percent error is less than 5\%  shown in Figure \ref{fig:6} contrasts starkly with the results for the candies-in-a-jar experiment (see Figure \ref{fig:3}) as now the group accuracy increases with increasing $N$. In addition,  there is a 30\% chance that a randomly selected  estimate  yields a prediction within the 5\% accuracy range, which certifies the competence of  the students to gauge lengths.

%----------------------------------------------------------------------------------------------------
%----------------------------------------------------------------------------------------------------

 The stark difference between  Figures \ref{fig:3} and \ref{fig:6} is somewhat reminiscent of  Condorcet's jury theorem (see, e.g., \cite{TJC}) in the sense that  adding more voters (i.e., increasing the number of estimates $N$)  may either improve or degrade the collective performance, depending on some circumstances. In Condorcet's  theorem, it is  the probability that  an individual  votes for the correct decision that determines whether a single voter or a jury will maximize the probability of making the correct decision.  In our case, this role is played by the percent error of the collective estimate: if it is too large, our results indicate that  discarding  individual estimates at random will increase the probability of  producing a high accuracy collective estimate.

\subsection{Estimating the number of pages of a book}

The same 139 students, who were asked to estimate the length of the paper strip, estimated also the number of pages of a book of $G=784$ pages. As in the previous experiment, there was  no reward or penalty for the subjects.  We recall that those students were very accurate on their estimates of the length of a paper strip (see Figure \ref{fig:4}). Surprisingly, their collective estimate of the number of pages of the book was kind of disastrous: $\langle g \rangle =  561$, which was superior to only 63\% of the individual estimates and corresponds to a percent error of  $28.4 \%$. The best guess was 800 pages that corresponds to a  percent error of  only $2 \%$.

 %----------------------------------------------------------------------------------------------------
\begin{figure}
\centering  
\includegraphics[width=0.48\textwidth]{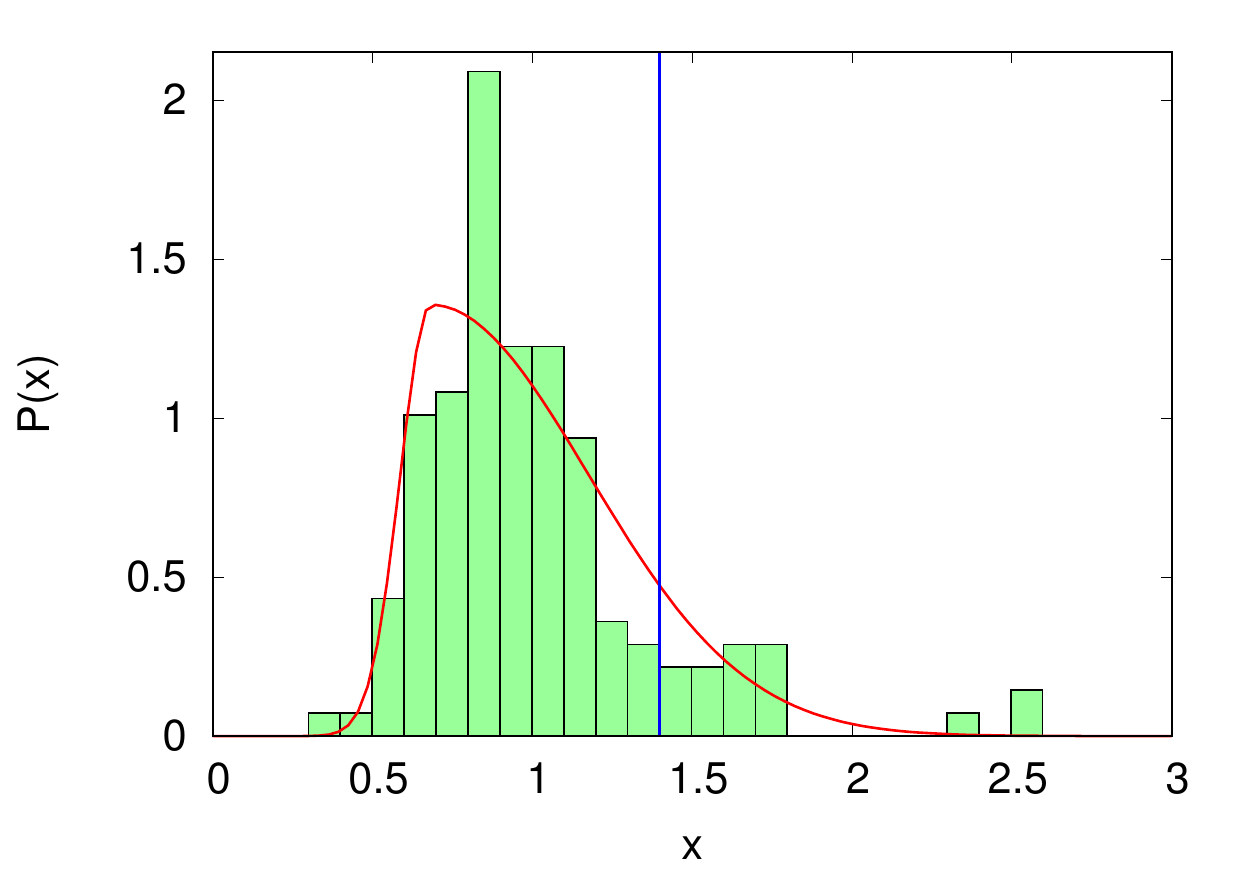} 
\caption{Histogram of the normalized estimates $x = g_i/ \langle g \rangle$ for the pages-of-a-book experiment. The vertical line indicates the ratio between the correct number of pages $G =784$  and the estimate of the group $\langle g \rangle =  561$, that is,  $G/\langle g \rangle \approx 1.4$. The solid curve is the best fitting ($R^2 = 0.84$) two-pieces normal distribution $ A \,  e^{-\left(x-\mu \right)^2/2\sigma_1^2}$  if $x < \mu$ and $ A \, e^{-\left(x-\mu \right)^2/2\sigma_2^2} $ otherwise, where $A = \left [ \sqrt{2 \pi} \left ( \sigma_1 + \sigma_2 \right )/2 \right ]^{-1}$
with   $\sigma_1 = 0.09$,  $\sigma_2 = 0.49$.  and $\mu = 1 - (\sigma_2 -\sigma_1) \sqrt{2/\pi} \approx  0.682$ so that the mean of the fitting distribution equals 1.  }
\label{fig:7}
\end{figure}
%----------------------------------------------------------------------------------------------------

Figure \ref{fig:7} shows the histogram of the  normalized guesses $g_i/\langle g \rangle $ which,  following Galton's  original hunch,  we attempted to fit using a two-pieces normal distribution  of mean 1. The histogram exhibits a considerable asymmetry that is measured by  the moment coefficient of skewness $\tilde{\mu}_3 = 1.22$. The analysis of the prediction power of random assembled combinations of $N$ estimates yields  results qualitatively similar to those of the candies-in-a-jar experiment. In particular, the scaled collective error is
$\gamma^{1/2}/G = 0.28$  and the scaled prediction diversity is  $\delta^{1/2}/\langle g \rangle = 0.36$. The pages-of-a-book experiment is an excellent illustration of the largely unsung but fundamental fact that the collective estimate may not produce accurate predictions.

\section{Discussion}\label{sec:disc}

The belief that cooperation can aid a group of agents to solve problems more efficiently than if those agents worked in isolation is a commonplace \cite{Huberman_90,Clearwater_91,Reia_19a}, although the factors that make cooperation effective still need much straightening out \cite{Reia_19b}. In fact,  this is the  main issue  addressed by the research on distributed cooperative or parallel problem solving systems \cite{Lazer_07}, since cooperation may well lead the group astray resulting in the so-called  madness of crowds as neatly expressed by MacKay almost two centuries ago: ``Men, it has been well said, think in herds; it will be seen that they go mad in herds, while they only recover their senses slowly, and one by one.'' \cite{MacKay_41}. 

However, the notion that  a  collection of independently deciding individuals is likely to predict better than individuals or even experts within the group  -- a phenomenon  dubbed wisdom of crowds \cite{Surowiecki_04} -- is much more controversial.  The first report of this phenomenon  in the literature was probably  Galton's  account of the surprisingly accurate estimate of the  weight of an ox  given by the median of the sample of the individual guesses \cite{Galton_07}.
% though the mean would  give an even   better estimate \cite{Wallis_14}.   
Although much of the evidence of the wisdom of crowds  is anecdotal (see, e.g., \cite{Surowiecki_04}), there are a few efforts aiming at  explaining this phenomenon  either using a purely statistical  framework \cite{Page_07} or using  psychological arguments on the nature of the individual estimates \cite{Nash_14,Nash_17}.

The main difficulty  to approach the wisdom of crowds phenomenon in a non-contentious manner is that it seems to have distinct meanings for  different researchers.  For instance, some  researchers view it as  the idea that a crowd can solve problems better than most individuals in it, including experts \cite{Mauboussin_07}. We note, however, that this is not what  Page's diversity prediction theorem asserts.
In fact,  equation (\ref{DPT})  asserts that $\gamma  \geq \epsilon $, where $\gamma $ is the quadratic collective error and $\epsilon$ is the average quadratic individual error, which equals the expected error of the estimate of a randomly selected  individual in the group.  Hence, the theorem offers  no guarantee that the collective estimate will be better than the estimates of  most individuals in the group. Nevertheless,
for the pages-of-a-book experiment we find that the collective estimate is better than 63\% of the individual estimates, whereas this figure increases to 85\% for the paper strip experiment.   We note that  the specification `including experts'  in the above definition is misleading   because one may be led to believe incorrectly that the collective estimate is better than the experts'. For example,   even in the paper strip experiment for which the collective estimate was highly accurate, it was outperformed by 15\% of the individual estimates. 

At this stage we should note that our disagreement over Page's diversity prediction theorem $\gamma =  \epsilon - \delta$ (see equation (\ref{DPT})), or, more correctly, over the interpretation of that theorem, since its proof is straightforward, is that one cannot  infer how  a change on the diversity of the estimates $\delta$  will influence the collective error $\gamma$, because the mean individual error $\epsilon$ will  also be affected by that change. Nevertheless, the effect of $\delta$ on $\gamma$ is an important  issue that can be addressed empirically. Since it is not feasible to carry out many independent experiments to calculate the correlation between these quantities,  here we produced  those experiments artificially by selecting $N$ estimates at random and without replacement from the original set of estimates of our experiments. Our results (see Figures \ref{fig:2} and \ref{fig:5}) indicate that there is no significant correlation between $\gamma$ and $\delta$, that is, diversity has no predictive value at all for the accuracy of the collective estimate.

We think the reason the phenomenon of the wisdom of crowds caught  on has little to do with the fact that on the average the collective estimate  improves upon  randomly chosen individual estimates, which can actually  be quite poor as illustrated by our pages-of-a-book experiment. For many researchers (see, e.g., \cite{Nash_17}), the real riddle is  the surprisingly good accuracy of the  collective estimate that,   in Galton's seminal experiment, missed the correct weight  by 0.8\% only \cite{Galton_07}.  A possible explanation involves  the combination of forecasts \cite{Bates_69} which, on the condition that the  individual forecasts are unbiased, guarantees that the accuracy of the combined estimation increases as the number of independent estimates increases.   The trouble with this approach is the implausibility of the assumption that  the individual estimates are unbiased, that is, that their means coincide with the correct value of the quantity being estimate.  (It is interesting that the assumption of unbiased individual estimates  means that  one could harvest the benefits of the  wisdom of crowds by asking a single individual to make several  estimates at different times \cite{Vul_08}.) 
On the contrary,  there seems to be a systematic error in the wisdom of crowds  so that the collective estimate depends on some  typical, group-dependent  belief on the value of the unknown quantity \cite{Nash_14,Nash_17}.  Since this typical value may be quite apart from the correct value, there is no guarantee of the accuracy of the collective estimate, which is precisely the conclusion we draw from the experiments reported here. Hence, the high accuracy of the  collective estimate, which gives  the wisdom of crowds its popular appeal,  is most likely an artifice of selective attention or cherry picking that gives prominence to the successful  outcomes only.

\bigskip

\bigskip

\acknowledgments
The research of JFF was  supported in part 
 by Grant No.\  2017/23288-0, Fun\-da\-\c{c}\~ao de Amparo \`a Pesquisa do Estado de S\~ao Paulo 
(FAPESP) and  by Grant No.\ 305058/2017-7, Conselho Nacional de Desenvolvimento 
Cient\'{\i}\-fi\-co e Tecnol\'ogico (CNPq).
DAN  was supported by grant  2017/08475-9, Fun\-da\-\c{c}\~ao de Amparo \`a Pesquisa do Estado de S\~ao Paulo 
(FAPESP).

\end{document}